# Experiential AI: A transdisciplinary framework for legibility and agency in AI

Drew Hemment[1*], Dave Murray-Rust[2], Vaishak Belle[3], Ruth Aylett[4], Matjaz Vidmar[5], Frank Broz[6]

[1] Edinburgh Futures Institute, Edinburgh College of Arts, University of Edinburgh, Edinburgh, Scotland
[2] Faculty of Industrial Design Engineering, TU Delft, Delft, Netherlands
[3] School of Informatics, University of Edinburgh, Edinburgh, Scotland
[4] School of Mathematics and Computer Science, Heriot-Watt University, Edinburgh, Scotland
[5] School of Engineering, University of Edinburgh, Edinburgh, Scotland
[6] Interactive Intelligence Group, TU Delft, Delft, Netherlands

# Abstract

Experiential AI is presented as a research agenda in which scientists and artists come together to investigate the entanglements between humans and machines, and an approach to human-machine learning and development where knowledge is created through the transformation of experience. The paper discusses advances and limitations in the field of explainable AI; the contribution the arts can offer to address those limitations; and methods to bring creative practice together with emerging technology to create rich experiences that shed light on novel socio-technical systems, changing the way that publics, scientists and practitioners think about AI.

# 1. Introduction

We are witnessing an explosion in data-driven artificial intelligence (AI), with profound implications for the creative industries and for society at large. This includes generative AI tools such as ChatGPT, DALL-E 2, Stable Diffusion and Midjourney, driven by recent developments in diffusion models and large language models (LLMs). Due to the move to data-driven systems and increased complexity of algorithms, these tools are inscrutable and opaque to human understanding, which limits scope for human intervention and agency, and presents the primary challenge to their usability, fairness and accountability. Everything from the way data is collected, the labelling and cleaning processes, and both training and testing regimes, profoundly determine the type of outputs and decisions generated, and their impacts on end-users. Earlier goal-driven systems could at least articulate in some form what their goal was, where data-driven approaches require a detailed understanding of the often dynamic data context in which they operate. Moreover powerful and widely applied algorithms such as deep learning encode the system's knowledge implicitly and in a distributed fashion so that even experts may not easily be able to determine what the system 'knows'.

Explainable AI (XAI) is the leading current approach within the AI community investigating how the decisions and actions of machines can be made explicable to human users (Gunning 2017; Belle & Papantonis 2021). Core to this is the notion that transparency not only requires that the operation of the system be visible, but that it can provide comprehensible explanations as well. Additionally, unless people can access comprehensible explanations, that system cannot be held to account. And for there to be fairness, explanations are needed to equip individuals to make good decisions about the use of a system, contest its outputs, or help shape its design and operation.

Here we report on work in arts, technology and futures towards an experiential approach to machine intelligence and user interaction to underpin new paradigms for legible and inclusive AI. We argue that despite significant achievements in XAI, the field is limited by a focus on technical explanations. Turning to the arts for inspiration, we propose that experiential methods can engage people emotionally, cognitively and tangibly with the large scale effects of pervasive AI deployments. Such experiences can enable a holistic engagement, that is particularly relevant to understand the nuanced social implications of deployed AI technology, alongside and not isolated from socio-technical applications. We argue an experiential approach can enhance and complement current work on explainability, by generating the kind of understanding needed from a human point of view, in particular for non-experts using or impacted by AI. Looking beyond explanation, experiential understanding can help to support other kinds of engagement, such as contestation around models and their design (Alfrink et al. 2022). This in turn increases scope for human agency, intervention and control.

A precursor to this work is Experiential AI (Hemment et al 2019) which gave the outline of the trans-disciplinary perspective that we articulate and develop in this paper. Here, we further develop Experiential AI as a practice that opens up the AI field to greater understanding and collaboration between human and machine. We present new methodology and report on our early stage explorations in Experiential AI, in which professional artists worked with scientists, engineers and applied ethicists to create near future scenarios in which ML systems, social robots or other technologies can be designed, deployed and tested as experiences, in the form of interfaces, installations, performances, situations, interactions. This paper explores how this practice offers new modalities of explanation, that open up algorithms, the science behind them, and their potential impacts in the world to scrutiny and debate. We discuss how such as approach can use art and tangible experiences to communicate black-boxed decisions and nuanced social implications in engaging, experiential ways, with high fidelity to the concepts. Moreover, beyond communicating current knowledge, we propose an experiential approach can generate new understanding on AI systems – their operations, limitations, peculiarities and implications. We reflect on how the methodology can be used to make the opaque mechanisms of AI artefacts and algorithms transparent to those who interact with them, in order to increase the range of people engaged in shaping the field and restore the basis for accountability. Going beyond explanation, the paper develops Experiential AI as an approach to human-machine learning and development where knowledge is created through the transformation of experience.

## 2. Recent advances and current limitations in explainability and explainable AI

There is not yet a shared definition of XAI within the AI community, but for the purposes of this paper, we take it to denote interpretable features within an algorithm that enable decisions to be justified, tracked and verified by a human (Montavon et al. 2018). Some strands of research focus on using simpler models (possibly at the cost of prediction accuracy), others attempt "local" explanations that identify interpretable patterns in regions of interest (Weld & Bansal 2018; Ribeiro et al. 2016), while still others attempt human-readable reconstructions of high-dimensional data (Penkov & Ramamoorthy 2017; Belle 2017).

While development in the field has been exciting, we identify the following six gaps or areas for development in the current XAI landscape, with increasing relevance to the Experiential AI field:

**1. Providing explanations from human points of view**
Such work often struggles to generate the kind of explanations needed from a human point of view. A person won't always need to know in detail how black-box AI works, or be helped by an opaque description of machine logic. They will instead want to understand their limitations, when their outputs can be trusted, or why there was one outcome and not another. Recent XAI work acknowledged the need to develop alternative, human-centred routes to understanding (Ehsan et al. 2022) through lenses like sensemaking (Kaur et al. 2022 ) and user experience (Liao & Varshney 2022). It is clear that seeing inside something is not the same as understanding it, and that developing understanding places substantial burdens on people to seek out information and make sense of systems (Ananny & Crawford, 2018), mutating social into personal responsibility.

**2. Looking beyond model explanations to address the entirety of the AI system**
There have been impressive advancements made in XAI, that include the socio-technical aspects of the system, including not just the models at their core, but the data collection and processing that gives rise to them and the way the system has been commissioned and designed (for example Ehsan et al. 2021). However, there remains an urgent need to understand the entirety of AI systems, including the relations between the system and the subjects of its operations. As now widely acknowledged, most machine learning systems cannot be disentangled from their data sources, raising privacy and ethical concerns, and leaving the user with the burden of understanding this.

**3. Connecting technical aspects to higher-level dimensions of AI**
The vast majority of advances in XAI focus on the technical operation of AI (Arrieta et al. 2019, Belle and Papantonis 2021), rather than it's embedding in society, the politics and sociology that surround it, and the effects that is has in the world. By and large, current work in XAI addresses explainability as primarily a lower-level technical problem, and does not account for the higher-level – system-level, cognitive, political, legal, regulatory or institutional – aspects of AI. There is a need "to consider how AI fits into the wider socio-technical context of its deployment" (Royal Society 2019), particularly in light of developments such as the EU AI Act that start to require explanation and transparency around models.

**4. Accounting for a wider range of stakeholders for systems deployed in social situations**
XAI has to date predominantly targeted a narrow range of expert users. As AI-enhanced applications are increasingly deployed and situated in social settings, actionable insight on both their operation and implications needs to be communicated to the people they impact. More intuitive solutions are required to, for example, integrate domain knowledge in ways that connect users to those impacted by the system (Veale et al. 2018) and that support engagement by a broad segment of society.

**5. Engaging with the imaginaries surrounding AI**
AI and robots are the subject of widespread illusions, for example that a machine 'belief state' is comparable to a human mental state. There are strong imaginaries around AI (Bory & Bory 2015), that can lead to both exaggerated fears, and an unrealistic fascination and over-imputation of authority (Robinette et al. 2016). Engaging with the imaginaries and misrepresentations that surround AI can help to better orient policy, development and public debate (Natale & Ballatore 2020).

**6. Deeper engagement in material and ideological realities**
In the round, it is not clear that abstract models can incorporate cultural and sociopolitical norms in a straightforward manner. Ananny & Crawford (2018) call for "a deeper engagement with the material and ideological realities of contemporary computation" (*Ibid.*). This requires moving beyond explanation to creating a deeper understanding of AI, in a manner that supports engagement by a broad segment of society and addresses the nuances of socio-technical systems.

# 3.  AI art and experiential learning

Research on experiential learning (Kolb 2014) confirms that artistic practice can act as a powerful pedagogic mechanism. A range of creative methods can be used to create experiential learning settings, including sensory and affective engagement to dramatise concepts, promote the freedom to act and explore the consequences, and generate reflection. This can include embedding relevant experiences in a story-world through narrative, or role-play, as for example Boals Forum Theatre (Boal 2013).

Experiential methods have been shown (Vannini et al. 2011) to work better than a purely knowledge-based or data-driven approach, and are known to be more effective than merely providing information or logical arguments – showing rather than telling, to create deeper understanding. Long-established theories of experiential knowledge describe it as is a function of the type of acquired information and one's attitude towards said information (Borkman 1976). The quotidian experiences of daily AI that is aware of humans (Kambhampati 2020) provide a rich ground for experiential learning. Examples from human-robot interaction demonstrate that models of how we interact with technology that don't take experience into account may be incomplete or incorrect (Smedegaard 2019).

The potential contribution of the arts to explainability remains largely untapped. There are promising developments around the use of generative artworks as educational tools to explore AI ethics (Srinivasan & Parikh 2021), and steps towards reframing explanation in a more holistic way, for example the Living Room of the Future (Sailaja et al. 2019) that makes experientially stark issues of agency, legibility, privacy and trust. Stephen Wilson (2002) identified characteristics of 'information arts' that offer a basis for *alternatives in setting research agendas, interpreting results, and communicating findings*:
"• A tradition of iconoclasm means that artists are likely to take up lines of inquiry devalued by others.

> • The valuing of social commentary means that artists are likely to integrate widely ranging cultural issues into their research.
> • Artists are more likely than commercial enterprises to incorporate criteria such as celebration and wonder.
> • Interest in communication means that artists could bring the scientific and technological possibilities to a wider public.
> • The valuing of creativity and innovation means that new perspectives might be applied to inquiries."
> (*Ibid*.: 38)

In a comprehensive review of contemporary AI art, Dejan Grba (2022) notes that "Digital technologies offer a generous space for conceptual, as well as formal, methodological, and aesthetic experimentation that can transcend the technologically imposed limits of expression." Grba argues the 'poetic' dimension of AI art is "informed by the various phenomenological aspects of sub-symbolic ML systems" (*Ibid*.), and is comprised of strategies that: "explore the epistemological boundaries and artifacts of ML architectures; sample the latent space of DL networks; aestheticize or spectacularize the renderings of ML data; and critique the conceptual, existential, or socio-political consequences of applied AI; a few works criticize AI art itself" (*Ibid*.). Grba is concerned not with art as experiential learning per se, but with the "integrity and inspiration of avant-garde art practices" (*Ibid*.). Nonetheless, in interrogating art practices as critique, he provides a compelling and rigorous account of their explanatory potential and limitations.

In our empirical Experiential AI research (see appendices), we conducted a co-operative research inquiry (Heron and Reason 2001) involving a multi-disciplinary team of professional artists and design, XAI and applied ethics scholars. Here, we describe three case studies arising through this work, and give a short summary of findings that relate to the potential for experiential learning:

| Project | Theme | Explanation |
|---|---|---|
| The Zizi Show by Jake Elwes (Supplement 1) | Bias in ML data, and misrepresentations of AI. | The Zizi Show generates imagery of non-binary bodies in order to bring attention to the underrepresentation of LGBTQ+ communities in ML training data. It is an explanation through experiential means of a dense clustering of issues: bias in ML, lack of representation, non-binary identities, and anthropomorphism in AI. Zizi highlights the way data and design choices shape what ML does. It shows how the model learns a representation of people, that is embedded social life. Elwes directly engaged and empowered an underrepresented community, providing paid employment, accreditation, and agency over the way data is stored. The project engages a marginalised group, and develops their literacy surrounding bias in ML, thereby supporting their agency in contesting its fairness and accountability. Zizi shows end users there is something to contest, even if that do not interact directly with the model themselves. Zizi specifically targets anthropomorphised misrepresentation of AI, by constructing an AI persona, and then deconstructing it, and exposing its construction in software by the human artist. By highlighting correspondences between AI and drag at a surface level, it asks deeper questions about the character of statistical knowledge applied to shifting human identities. |
| Cypress Trees by Anna Ridler and Caroline Sinders (Supplement 2) | Hidden human labour in ML. | The project highlights that what we think of as ML intelligence is actually human intelligence at many points in the system. It explains the labour in ML systems, often performed by so-called click workers, or the call centre workers having to listen to user responses. The way the artists build their own data in a very meticulous and painstaking way points to wider issues of hidden human labour in ML data. The artists in this sense 'perform' a part of the machine learning algorithm. They turn a foundational definition in AI on its head, by the human artist doing a task usually done by the computer and associated with machine intelligence. Rather than for problem-solving, the human-machine intelligence is applied to produce imagery and gallery installations that represent the ordering of knowledge by AI. The work debunks the neat representations of 'autonomous' systems, and by forcing an attention to hidden labour, raises wider questions around human bias and worker exploitation. |

| The New Real Observatory by The New Real and Various Artists (Supplement 3) | Increase legibility through accessible tools for co-creation and collaboration with AI. | The AI platform opens data and algorithms up for exploration and discovery by the artists, and enables reflection on novel concepts for human-AI co-creation. The artworks by Inés Cámara Leret, Adam Harvey, and Keziah MacNeill each offer a specific insight, perspective, dimension, or lens through which AI can be interpreted, and in different ways provide 'ground truth' to computational ways of making sense of the world. Cámara Leret is interested in the impact of AI in both enabling and hindering our understanding of, and relationship with, the environment. In The Overlay, she explores the entanglements that arise when translating global data to local environments. In his artwork and an accompanying essay, Harvey questions the way GANs and generative AI automate artistic production and auto-complete visual concepts, transforming low-dimensional ideas into high-resolution imagery, while highlighting concerns ranging from computational plagiarism to excessive energy usage. MacNeill creates a dialogue between artificial intelligence, climate data and photography, tuning into the operation of the neural network to stimulate novel thinking on the character of the algorithmically generated image. Wrap-around activity including talks and workshops enabled the audience to explore how creative interventions can critically interrogate AI technologies and fuel transformative experiences for audiences. |
|---|---|---|

We identify in our case studies the following contributions to the aforementioned limitations in XAI:

**Gap 1. Providing explanations from human points of view**
The arts develop symbolic representations and narratives alongside statistical information, and artists are adept at surfacing issues or ambiguities that will benefit from explanation. Imagery, narrative and interaction are used to probe, explore and communicate significant aspects of technology in imaginative and absorbing ways, engaging people emotionally, cognitively and tangibly. The artworks can provide an account of why the AI did one thing and not something else. Yet they are not didactic explanations, but concrete artefacts or representations that can be used to illustrate attributes and concepts, and as an object to spark further enquiry and explore deeper meaning.

**Gap 2. Looking beyond model explanations to address the entirety of the AI system**
The case studies go beyond a narrow account of model interpretability to address different aspects of the ML pipeline and issues such as human labour in automated systems. Such works consider the entirety of AI systems and the operation and implications of AI in real-world settings for human experience and relationships. In each case, the 'explanation' is not a technical account of the model or algorithm, so much as a creative representation of an AI system situated in the world, and its attributes or implications.

**Gap 3. Connecting technical aspects to higher-level dimensions of AI**
There is a high level aspect of the ML system that is being made explicit and communicated in each case, such as bias and inclusion in ML data. These art projects make connections beyond the models, algorithms and datasets out to their social contexts and implications. The artistic experiments bring to life and question not only what an algorithm does, but also what a system could be used for, and who is in control.

**Gap 4. Accounting for a wider range of stakeholders for systems deployed in social situations**
The AI art projects illustrate how to improve understanding for non-experts who are not supported by mainstream XAI. They engage demographic groups who are underrepresented in training data and in the design and evaluation of these systems. To this extent, they suggest strategies that can help to make AI more accessible and help to reach those currently excluded from the creation and deployment of systems.

**Gap 5. Engaging with the imaginaries surrounding AI**
The artworks provide speculation about the future capabilities of ML models – what happens when they start to reproduce bodies, movements and other parts of cultural ways of being. We see here artists seeking to demystify AI, by addressing issues including anthropomorphism and exaggerated accounts of the autonomy of AI systems. This said, it is not hard to also find instances of artistic practice that serve to obscure rather than articulate, or amplify a misconception rather than demystify.

**Gap 6. Deeper engagement in material and ideological realities**

Such works enable the artifacts of machine reasoning to be made tangible, and enable complex philosophical, political, moral and technical questions to be explored experientially and when embedded in real situations. They offer an exploratory rather than didactic mode of 'explanation', which can advance cross-disciplinary understanding.

# 4. Towards an approach and methodology for experiential AI to enhance legibility and agency

We see here ways in which artistic practice can enhance the legibility of intelligent systems and scaffold understanding of AI. And yet, the arts are not instruction, and it would be wrong to instrumentalise the arts, either for science communication or system design.

Our preliminary audience evaluation of the case studies indicates that learning outcomes rarely follow from an artifact or performance alone. The artists consider their works to be not just the artifact or performance, but also the accompanying talks, articles and workshops, and we found the wider temporal envelope and wrap-around engagement is key to the learning outcomes for an audience. This leads us to a vision for experiential learning in intelligent systems where an explanation is not a thing a model produces, but a learning journey, a process and an interaction between the AI and the user (O'Hara 2020).

Grba highlights both the power and limitations of contemporary AI art as critique, and argues "the uneven intellectual breadth and depth, biased or constrained contextual awareness, and sketchy art-historical knowledge affect many AI artists' conceptual thinking, methodologies, and the cogency of their outcomes" (2002).

Therefore, we develop Experiential AI as a cross- or trans-disciplinary methodology, that is complementary and additional to the practice of artists, and that has a specific focus on orchestrating and brokering between disciplines. This builds on the Open Prototyping framework (reported in Leonardo in Hemment 2020), and has the following objectives:
1. Connect the technology to social goals;
2. Combine the efforts of artists and XAI practitioners, while preserving an autonomy for the practice of individual contributors;
3. Ground the art in current science;
4. Develop creative expressions with high fidelity to the technical concepts;
5. Provide an interface to the public and policy debates;
6. Enable rigorous evaluation.

Elsewhere, we have proposed that Experiential AI can be characterised by four dimensions which we refer to as the 4As conceptual model: Aspect, the socio-cultural and institutional dimension; Algorithm, the system and technology design; Affect, the experience and its reception; Apprehension, its impact on an audience (compare Hemment et al 2022; Hemment et al Forthcoming).

Here we construct a 'process model' for Experiential AI (Figure 1) in the tradition of the Design Council's Double Diamond (2005), by mapping these four dimensions onto the Open Prototyping process model (*Ibid*.).

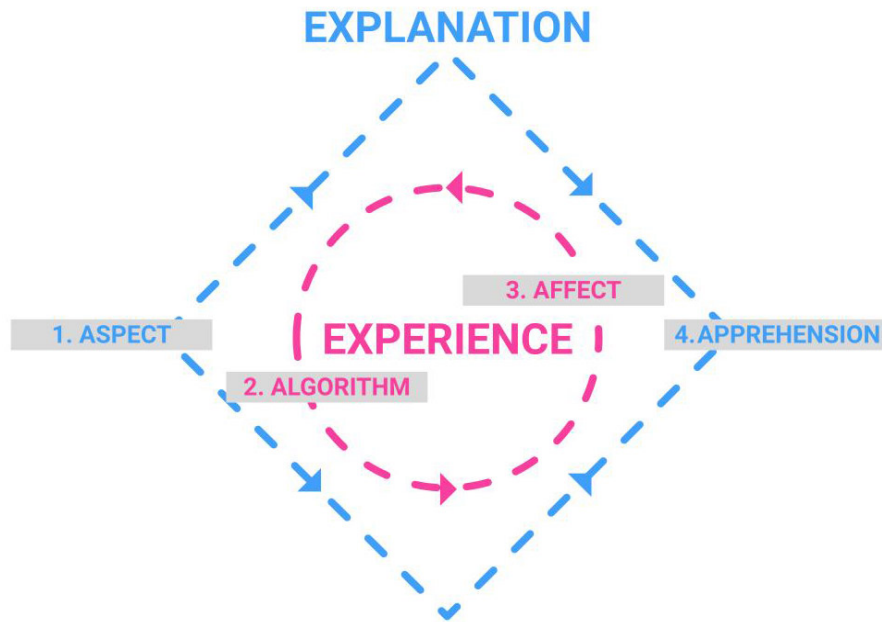

**Figure 1: Experiential AI Process Model**

The process model, then, provides a scaffold for encounters and experiences to generate cognitive engagement and learning. We envision multi-disciplinary teams, that come together to scope issues of concern and future scenarios (Aspect), and in which XAI practitioners articulate definitions of interpretable aspects and features (Algorithm), and artists conduct creative experiments with AI and emerging technology that generate sensory experiences (Affect), so that technical systems and creative works are built that enact or implement insights on transparent and responsible AI, with which an audience interacts to generate learning outcomes (Apprehension).

The result is a tool and method that can be used to design and evaluate creative experiments in which AI researchers and artists are jointly engaged to make data-driven AI and machine learning tangible, interpretable, and accessible to the intervention of a user or audience. We can think of the diagram as a gear wheel that connects machine intelligence to human learning, and technical features to higher level aspects. It can be used to represent the interactivity between, on the one hand, the machine intelligence that extracts patterns from observed data to generate decisions and outputs, and, on the other, human judgement, creativity and comprehension. It provides a model for human-machine learning and development where knowledge is created through the transformation of experience. The explanation is baked into the design of the experience, either through the curation of data, the design of the algorithm, or the way the components are connected.

This method can be applied to develop experiences with explanatory skill for various aspects of the life cycle of AI systems, from data collection, systems design, algorithm selection and deployment, through to the interests and ideologies vested in their decisions and the social implications that follow. This includes not just the models at their core, but also the data collection and processing that gives rise to them, the way the system has been commissioned and designed, how automated decision making is situated, what the model actually does, as well as the relations between the system and the subjects of its decisions.

This in turn, can feed back into the design of technologies, shaping XAI development. Using the tool, designers and users of intelligent systems can create scenarios in which a higher-level aspects or assumption can be communicated to a user, then develop the architecture and algorithm, and integrate data and model, in a tangible output that makes the aspect explicit, that users engage with to derive a demonstrable learning outcome. This involves using experiential methods to explore AI technology in concert with publics, and drawing on XAI philosophies, but going beyond functional explanation to develop contexts and implications for AI systems.

Adoption of this cross- or trans-disciplinary approach requires engagement between communities with different logics and rigours, different vocabularies and goals. The tool is not intended to take the place of a practitioner's own methodology, in XAI or the arts, but rather to facilitate and broker contributions from across different disciplines, or for an individual practitioner to guide a team through a trans-disciplinary process.

This work is ongoing and our future research aims to better understand the potential contribution of experiential explanations, by evaluating cognitive shifts and learning outcomes in users through audience studies, the significance of the works through art criticism, and the validity of the algorithms through quantitative experiments. One interest is to understand reinterpretations that can be made with the input of artists, while staying faithful to the AI models. Such research can evaluate how experiences may produce shifts in understanding for the public, increased agency for users, and also, potentially, actionable insights for AI researchers and practitioners.

# 5. Conclusion: The implications for legibility and agency in AI systems

We saw in the AI art projects a rich, multifaceted engagement in AI that has much to offer to science and society. Our reflection on the potential contributions of the arts to XAI has led to our central theme, namely, how artists, scientists and other interdisciplinary actors can come together to understand and communicate the functionality of AI, ML and intelligent robots, their limitations, and consequences, through informative and compelling experiences.

The outcome is Experiential AI as an approach to understanding AI systems by reconfiguring data, algorithms, models, interfaces and situations as experiences. Experiential AI provides a methodology for the arts and tangible experiences to mediate between impenetrable computer code and human understanding, making not just AI systems but also their values and implications more transparent, and therefore accountable. Moreover, it enhances and widens the scope of current work on explainability by engaging people emotionally, cognitively and tangibly with the effects of AI. This approach can open up algorithms, the science behind them, and their potential impacts in the world to public scrutiny and policy debate. Holistic questions can then be asked of the entanglements of humans and machines, going beyond model interpretability, such as how does AI challenge our world view, how does it shape human experience and relationships, and how we can avoid anthropomorphism and misplaced trust in AI. This can complement existing work in XAI that traces details of an algorithm's internal operation, by making tangible and illuminating underlying assumptions of machine learning models, processes that generate their data, or the social context in which automated decision making is situated. An experiential approach to AI can be used to help interacting humans to viscerally understand the complex causal chains in environments with data-driven AI components, including questions about: what data is collected, its nature, accuracy and freedom from bias, as well as who collects it; how the algorithms are chosen, commissioned and configured; and how humans are conditioned by their participation in algorithmic processes. Such experiential interventions can work to reach new audiences, to increase the agency of people impacted by these systems, and to create spaces for debate and engagement with populations outside the technical centre.

In this paper we have focused on XAI and the implications for explainability in AI systems. We further envision more widespread applications of the Experiential AI methodology, as an accessible, inclusive and intuitive approach to machine intelligence and user interaction. In this broader light, the field of Experiential AI seeks to engage practitioners in data science, robotics, AI art, design research and applied ethics, around an exploration of how humans and AI artefacts relate, through the physical and digital worlds, through decisions and shaping behaviour, through collaboration and co-creation, through intervening in existing situations and through creating new configurations. This offers a holistic and trans-disciplinary orientation in AI design, towards navigating profound challenges, exploration, and building agency.

This can have reciprocal benefits beyond the disciplinary perspectives. We hope this paper will contribute to research and practice on explainability, with a particular focus on supporting non-expect users and broadening the scope of current work to address the higher levels of explanation, understanding and context. Looking beyond explanation and XAI, we envision future research and development on creative strategies and methods which can add to the toolbox for human-centred AI, to generate inclusive and holistic understanding and more responsible AI design. The arts, meanwhile, are connected to the science and technology through Experiential AI, and, in turn, produce situated, embodied and intuitive meaning around algorithms and the effects of their deployments. We hope that the kinds of structured engagements set up through Experiential AI can help artists who are well versed in current technologies to benefit from emerging AI paradigms. We also hope that the depth and rigour of AI practice in the wider arts and creative sectors can be enhanced by enabling structured access to the significant science and through further contributions to the ways in which AI can be understood. More broadly, as well as spotting problems or limitations with current concepts, experiential approaches can contribute to greater understanding and collaboration between human and machine, and to public and policy debates around critical issues. The ambition is to help us think differently about how algorithms should be designed, and open possibilities for radically new concepts and paradigms.

**Appendix 1**
**Jake Elwes, The Zizi Show, 2020**

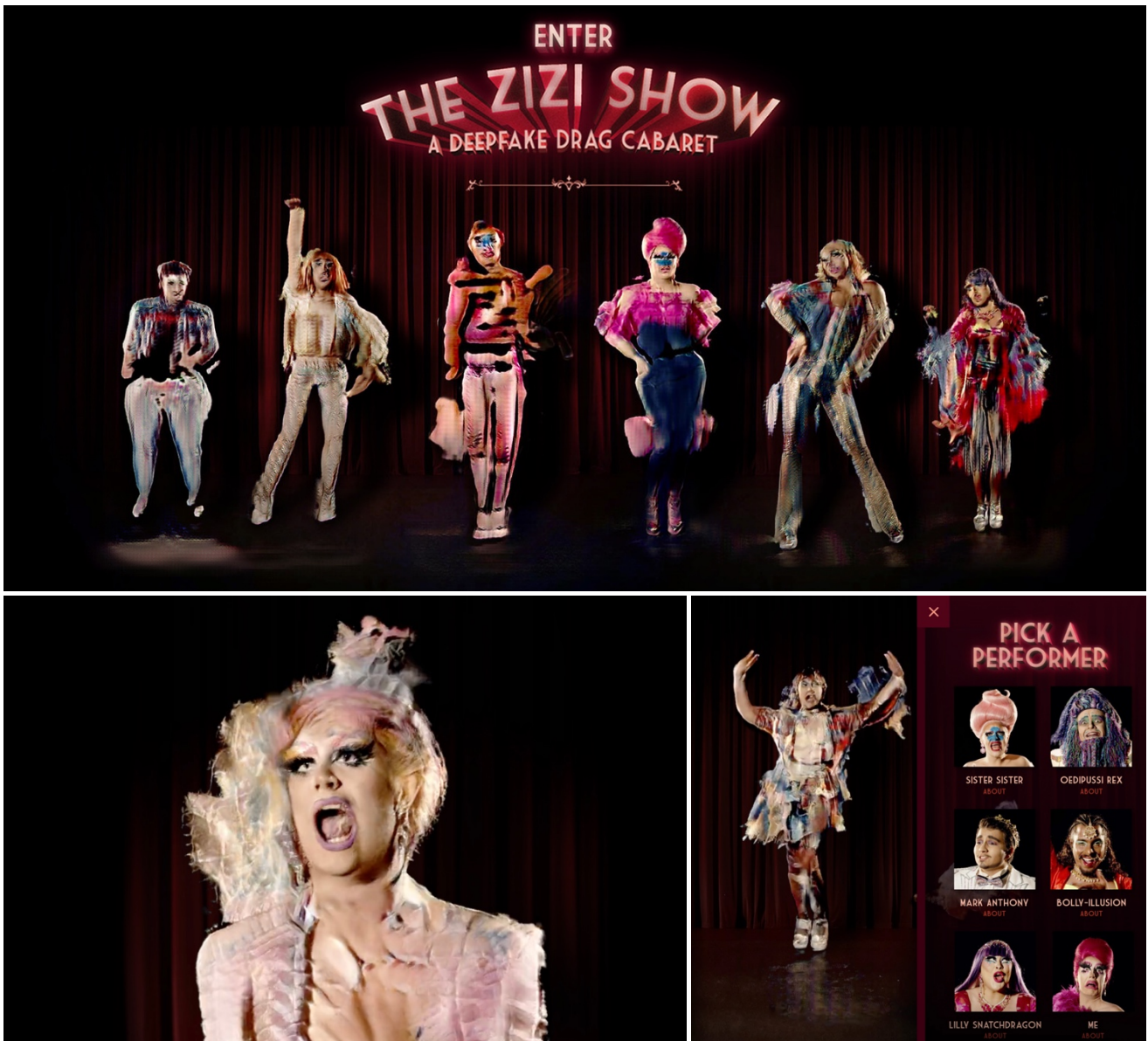

Figure 2 - The Zizi Show (Jake Elwes, https://zizi.ai/). An algorithmically generated compere asks the audience to select performers and songs. Each performer has a body blended from video capture of drag performers that morphs and changes as they perform each work. All Rights Reserved © Jake Elwes 2021.

The Zizi Show (Figure 2), by London-based visual artist Jake Elwes, is an online interactive artwork in which a GAN has been trained on video footage of thirteen diverse 'drag' performers, filmed at a London cabaret venue during the COVID19 lockdown. This work, builds on previous work (simply titled Zizi) that exposes the latent space of the ML model, and highlights the way the model outputs are shaped by the training data. Where many generative works have been trained on opportunistically collected data, the purposeful curation of Zizi's dataset explores the question of how human identity is represented within complex models. The Zizi Show develops this through digital avatars, that have been learned from real performers to create an interactive work that allows user control. A south London community of drag artists were engaged throughout, providing them with positive representation, safe spaces (an in-person venue and a secure server), and paid employment, during Covid lockdowns. Significantly, it connects low level technology to high level, social, cultural and political aspects of AI, such as ideas of cultural appropriation and machine bodies. It exposes the limits to machine intelligence, and inverts what is otherwise a deficiency in the technology, through a positive use of deep fake technology, in which a marginal identity is celebrated and embellished, rather than obscured or misrepresented.

1.1 Theme:
Bias in ML data, and misrepresentations of AI.

1.2 Intelligence:
The project engages with the current wave of machine learning techniques, using a StyleGAN network architecture re-trained on a modified version of Flickr-Faces-HQ (FFHQ) dataset, to which an additional 1,000 portraits were added, alongside a custom video, sound and interactive web interface. Machine learning here allows the creation of a generative space that includes bodies and faces.

1.3 Interactivity:
Zizi rethinks what interactivity is, at scale, and enables us to ask what forms of interactivity are important. In Zizi, the artist interacts with the model by manipulating data and weightings. The audience do not interact with the model directly, but with its artefacts and outputs. Moreover, they are able to do so at scale, as it has been designed for large numbers of simultaneous users. The audience view the output in different settings, and are able to select from a menu to switch between AI generated personas of drag artists for different music tracks. This, however, is representative of the interactivity experienced by a majority of users of intelligent systems. Many end users access the outputs of AI systems through interfaces that are not themselves intelligent.

1.4 Explanation:
The Zizi Show generates imagery of non-binary bodies in order to bring attention to the underrepresentation of LGBTQ+ communities in ML training data. It is an explanation through experiential means of a dense clustering of issues: bias in ML, lack of representation, non-binary identities, and anthropomorphism in AI. Zizi highlights the way data and design choices shape what ML does. It shows how the model learns a representation of people, that is embedded social life. Elwes directly engaged and empowered an underrepresented community, providing paid employment, accreditation, and agency over the way data is stored. The project engages a marginalised group, and develops their literacy surrounding bias in ML, thereby supporting their agency in contesting its fairness and accountability. Zizi shows end users there is something to contest, even if that do not interact directly with the model themselves. Zizi specifically targets anthropomorphised misrepresentation of AI, by constructing an AI persona, and then deconstructing it, and exposing its construction in software by the human artist. By highlighting correspondences between AI and drag at a surface level, it asks deeper questions about the character of statistical knowledge applied to shifting human identities.

**Appendix 2**
**Cypress Trees by Anna Ridler and Caroline Sinders**

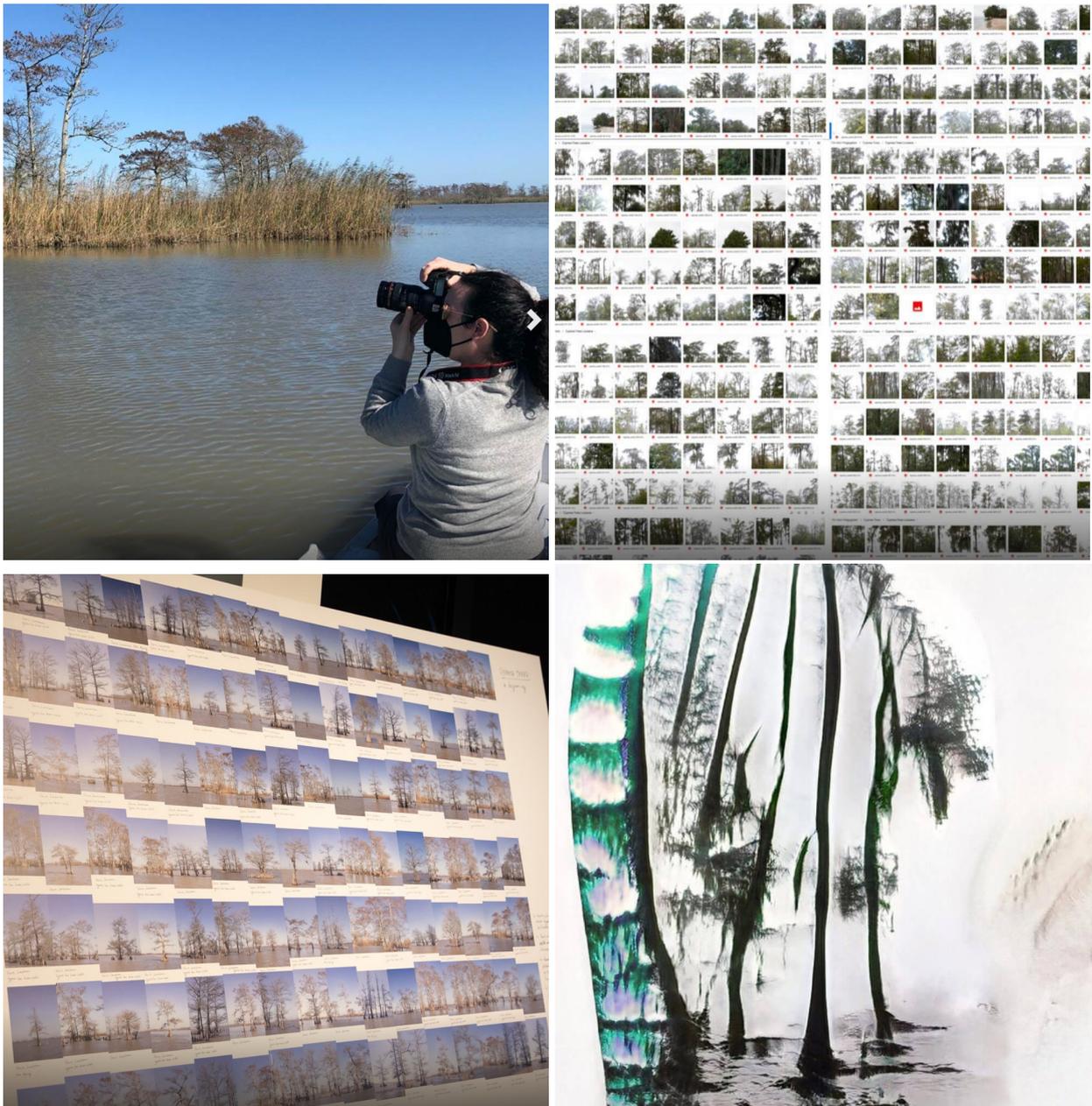

Figure 3 - Cypress Trees (Anna Ridler and Caroline Sinders, https://ars.electronica.art/newdigitaldeal/en/cypress-trees/), showing data collection (top left), data curation (top right), dataset display (bottom left) and initial experiements with GAN output (bottom right). All Rights Reserved © Anna Ridler & Caroline Sinders 2021.

How can AI help us to face the climate crisis and other entwined challenges? This machine-learning generated moving image piece (Figure 3) gives insights into the complexity of data sets and raises questions about deforestation and the politics of climate change, memory and loss. Anna Ridler and Caroline Sinders created a special dataset of the Bald Cypress on the gulf coast of the USA, where both have family ties. These trees, which can live thousands of years, are currently considered to be "threatened" by climate change. The work entails imagery of cypress trees produced by generative networks. At the heart of the work is the collection and curation of the datasets that the models require, drawing on the situational, embodied nature of machine learning systems.

2.1 Theme:
Hidden human labour in ML

2.2 Intelligence:
The artists painstakingly and meticulously extract patterns from observed data using manual methods in order to make judgements. The piece involves a generative network that produces images of cypress trees. However, the heart of the piece is the collection and curation of the datasets that this model requires, drawing on the situational, embodied nature of machine learning systems.

2.3 Interactivity:
It foregrounds the painstaking work by the artists to develop a bespoke data, through photography of hard to access trees, to labelling and cataloguing. The 'interaction' for an audience is to observe and move between artifacts the artist has curated and positioned in a gallery.

2.4 Explanation:
The project highlights that what we think of as ML intelligence is actually human intelligence at many points in the system. It explains the labour in ML systems, often performed by so-called click workers, or the call centre workers having to listen to user responses. The way the artists build their own data in a very meticulous and painstaking way points to wider issues of hidden human labour in ML data. The artists in this sense 'perform' a part of the machine learning algorithm. They turn a foundational definition in AI on its head, by the human artist doing a task usually done by the computer and associated with machine intelligence. Rather than for problem-solving, the human-machine intelligence is applied to produce imagery and gallery installations that represent the ordering of knowledge by AI. The work debunks the neat representations of 'autonomous' systems, and by forcing an attention to hidden labour, raises wider questions around human bias and worker exploitation.

**Appendix 3**
**The New Real Observatory**

The New Real Observatory is a creative AI platform developed with and for artists to enable multi-sensory exploration of possible futures and investigate the entanglements of people, data, machines and environments. The platform is powered by a series of available AI tools and processes, that have been integrated to allow users to articulate a dimension of change they wish to explore - either in visual (images) or symbolic (text) language. Artists Inés Cámara Leret, Adam Harvey, and Keziah MacNeill contributed to its development, and developed artistic prototypes that premiered at The New Real Pavilion at Ars Electronica 2022 in Linz, Austria. The artworks enable audiences to tangibly experience and explore the construction and artificiality of localised representations of nature generated by AI; the agency and control inherent in certain ways of creating and encoding data; and the speculative point at which the boundaries between humans, machines and nature blur.

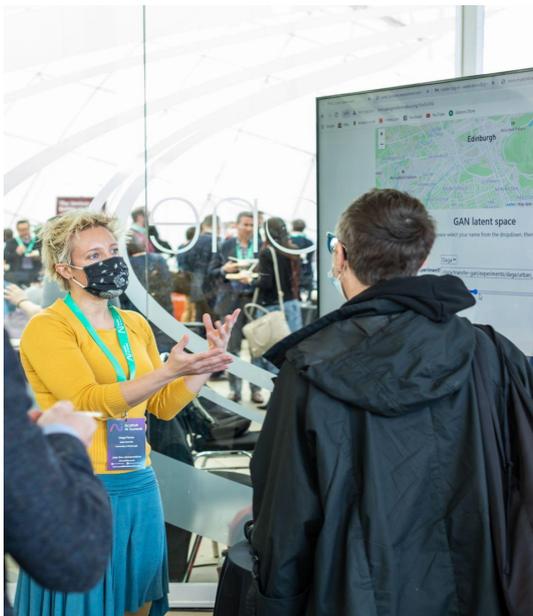
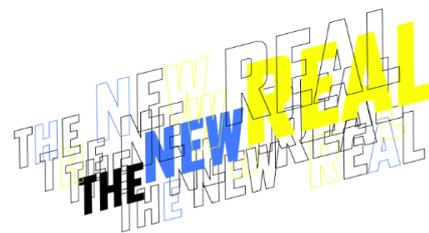
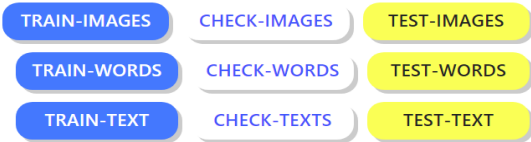

Image credit: *The New Real Observatory* creative AI platform. Image by The New Real, 2022.

**The Overlay**
**Inés Cámara Leret (ES)**
**2022**

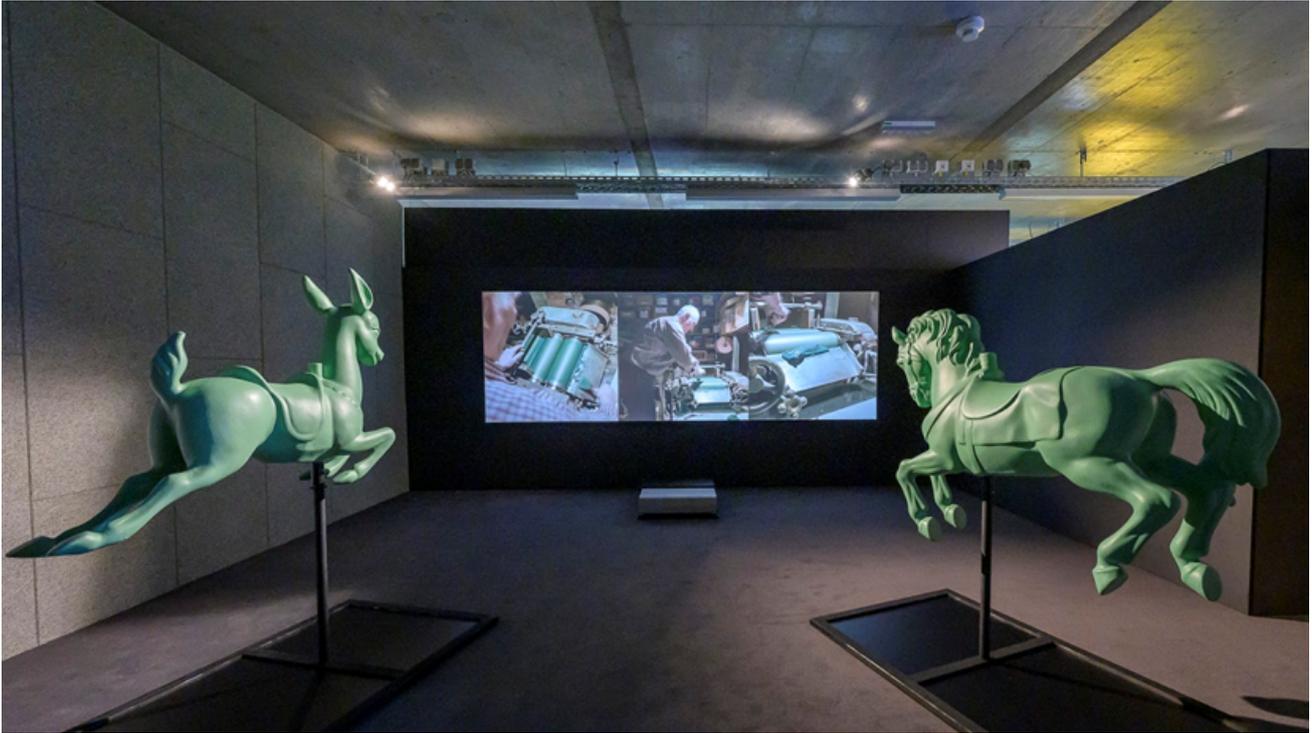

Image Credit: *The Overlay* by Inés Cámara Leret, 2022. Photo by DieFotoFrau.

The Overlay explores the construction and artificiality of localised representations of nature. The work references Disney's "go away green": a colour engineered to hide unsightly yet necessary objects in theme parks. Cámara Leret has collaborated with an AI processing engine fine-tuned on images of greenery and built environments to reveal hues of green for local neighbourhoods. The multi-component artwork features a digital interface that allows anyone to find their localised green for anywhere on Earth, recoloured objects from a traditional fairground ride, and a multi-channel film in which a local hue is fabricated by Spain's last living colourist and assimilated by the local community in its material form.

**Circular Diffusion**
**Adam Harvey (US/DE)**
**2022**

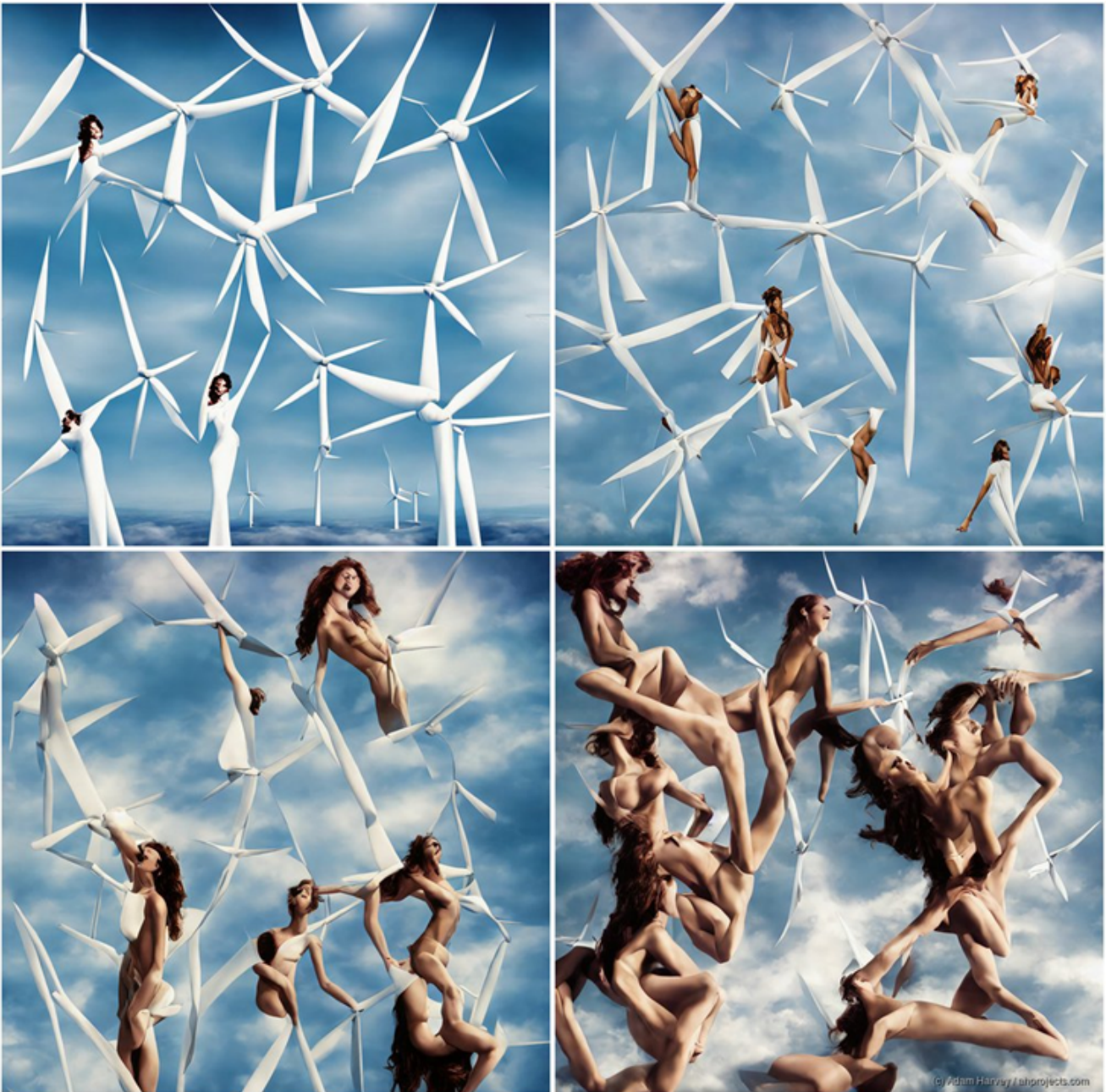

Image credit: *Sublime Diffusion N°2-5* © Adam Harvey, 2022.

In this new work Adam Harvey reflects on the perils and possibilities of generative AI technologies and their unavoidable relationship to energy and propaganda. The collection of images, titled "Circular Diffusion", references newly developed AI diffusion algorithms, their power to automate the production of awe-inspiring imagery, and the circular logic of extrapolation. AI is often considered a hopeful technology with unlimited problem solving capabilities. But new solutions create new problems. When GANs and Generative AI are applied to climate change they produce non-scientific output cloaked in scientific language. Further, using generative AI to address climate change can multiply the existing problem: mitigating climate change means reducing energy, but developing AI requires vast amounts of it.

**Photographic Cues**
**Keziah MacNeill (GB)**
**2022**

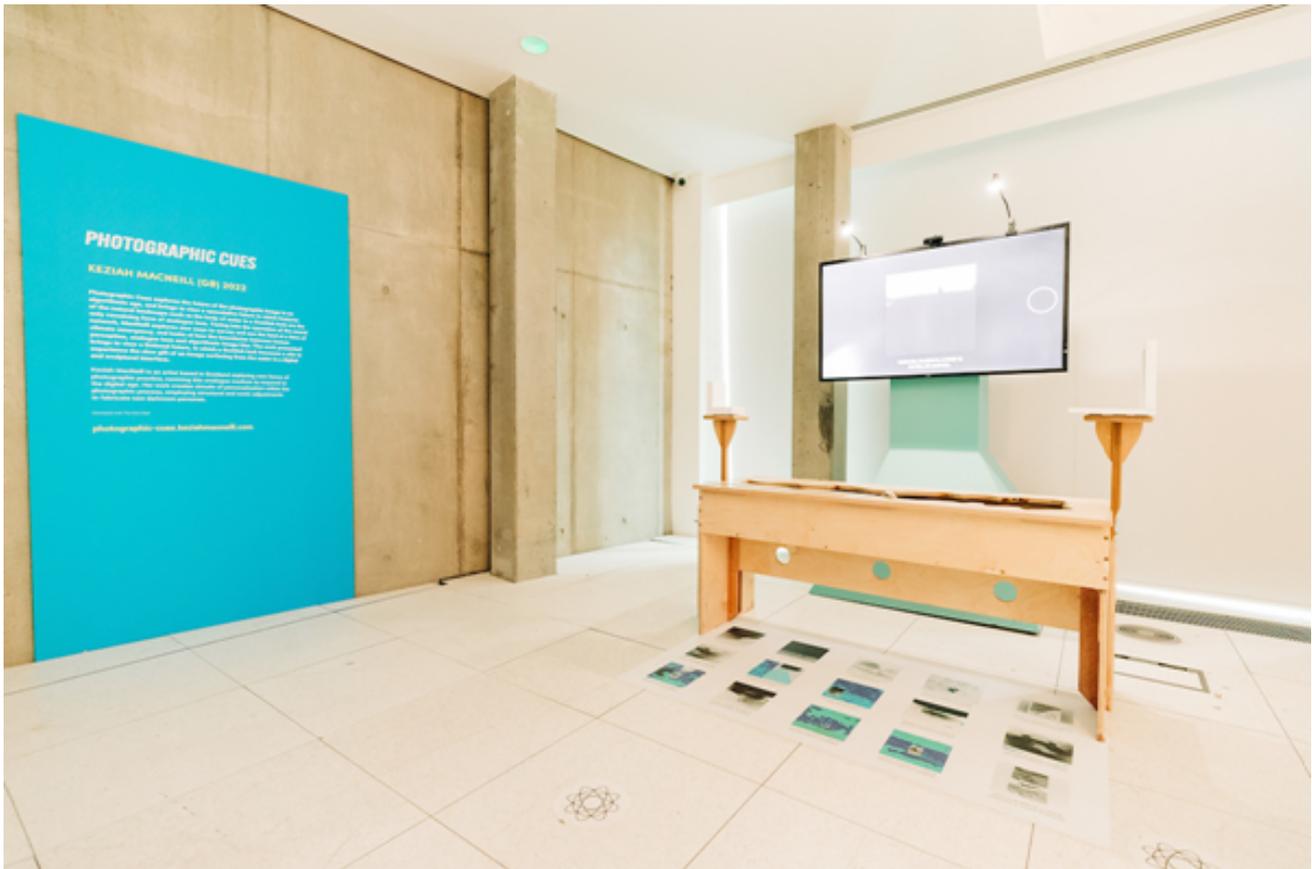

Image credit: *Photographic Cues* by Keziah MacNeill, 2022. Photo by Erika Stevenson.

Photographic Cues explores the future of the photographic image in an algorithmic age, and brings to view a speculative future in which features of the natural landscape such as the body of water in a Scottish loch are the only remaining form of analogue lens. Tuning into the operation of the neural network, MacNeill explores new ways to survey and see the land at a time of climate emergency. The work presented brings to view a fictional future, in which a Scottish loch becomes a site to experience the slow gift of an image surfacing from the water in a digital and sculptural interface.

**The New Real Pavilion**
**The New Real (GB)**
**2022**

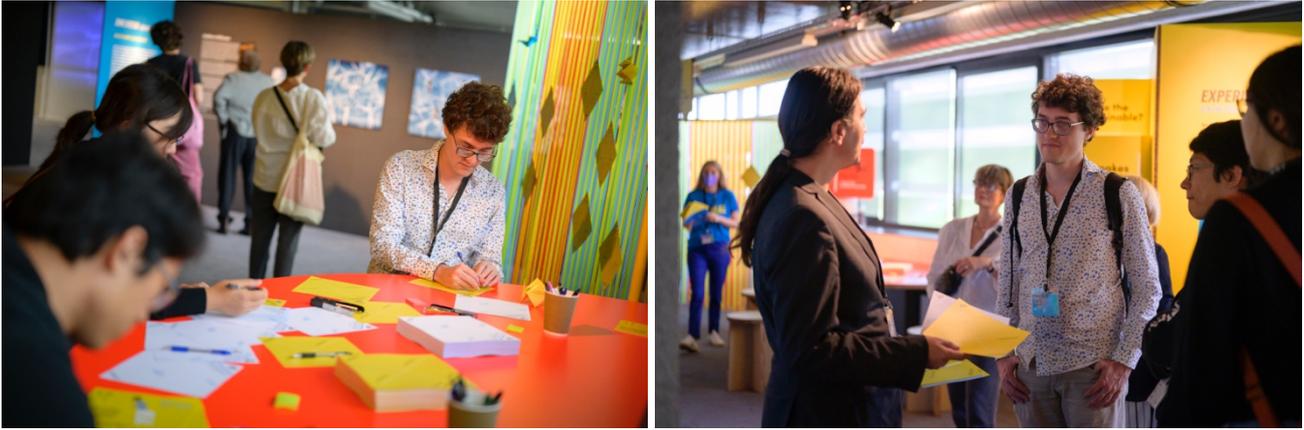

Image credit: *The New Real Pavilion and Research Hub*, The New Real, 2022. Photo by DieFotoFrau.

The artworks and a demo of the platform were contextualised by a pop-up research hub at The New Real Pavilion. Visitors to Ars Electronica 2022 were invited to talks and tours by artists, curators and scientists to create dialogue on the themes of the exhibition and artworks, and workshops and printed cards in the exhibition space elicited feedback and insights from festival attendees.

3.1 Theme:
Increase legibility through accessible tools for co-creation and collaboration with AI

3.2 Intelligence:
The platform is configured to allow artistic intervention through data curation, making explicit the fact that no machine can create ad novo in isolation. Through selecting data inputs to customise the processing engine, the artists articulate a conceptual dimension of change they wish to craft within the internal representation of the model. Once mapped, the dimension can then be explored to examine how a machine interpolates along a human-specified axis of meaning, enabling scaling with three key climate parameters: temperature, precipitation and wind speed at any point on Earth. The artworks combine the statistical reasoning of the ML networks with further mathematical calculation and her own artistic reasoning. This enables the artists to create alternative modalities of possible futures, by reinterpreting climate change data through imaginative recasting. The New Real Observatory highlights machine intelligence outside the current trends of generative networks and deep learning, specifically looking at planetary scale AI considerations (Bratton, 2016). Rather than centring on a single model, here predictive modelling is combined with an ecology of data flows, and then brought together around small interactive AI processes that produce images, words, numbers and sounds as responses to future possibilities and user input.

3.3 Interactivity:
The New Real Observatory features an online platform combining a database of localised climate modelling predictions, access to advanced Artificial Intelligence models and Machine Learning algorithms, and a powerful computational resource to allow exploratory experiments in algorithmic creativity, including a novel SLIDER tool to directly manipulate ML models. The design concept was to provide accessible tools for creative and ethical practitioners to probe a model or algorithm in order to conduct experiments and go beyond preset AI tools. The artists are able to interact with the model by, firstly, changing a future climate parameter that controls one of the weights for the model by specifying location and date, and, secondly, by curating and annotating data on which the model is trained. The platform has been used in five pilot projects with artists, and in future will be developed as a plug-and-play tool. In future iterations, audiences will be able to interact with the model by matching real-time (or past) GPS tracking with climate data-points, and by submitting new images, which are brought into the functioning of the piece. The platform generates visual, syntactic, audio or numerical outputs that the artists can use as material for art pieces, and, by extension, experiential explanations.

3.4 Explanation:
The AI platform opens data and algorithms up for exploration and discovery by the artists, and enables reflection on novel concepts for human-AI co-creation. The artworks by Inés Cámara Leret, Adam Harvey, and Keziah MacNeill each offer a specific insight, perspective, dimension, or lens through which AI can be interpreted, and in different ways provide 'ground truth' to computational ways of making sense of the world. Cámara Leret is interested in the impact of AI in both enabling and hindering our understanding of, and relationship with, the environment. In The Overlay, she explores the entanglements that arise when translating global data to local environments. In his artwork and an accompanying essay, Harvey questions the way GANs and generative AI automate artistic production and auto-complete visual concepts, transforming low-dimensional ideas into high-resolution imagery, while highlighting concerns ranging from computational plagiarism to excessive energy usage. MacNeill creates a dialogue between artificial intelligence, climate data and photography, tuning into the operation of the neural network to stimulate novel thinking on the character of the algorithmically generated image. Wrap-around activity including talks and workshops enabled the audience to explore how creative interventions can critically interrogate AI technologies and fuel transformative experiences for audiences.